\documentclass[aps,prl,twocolumn,superscriptaddress,unsortedaddress, footinbib]{revtex4}
\usepackage{graphics}
\usepackage[dvips]{graphicx}
\usepackage{amsmath}
\PassOptionsToPackage{caption=false}{subfig} 
\usepackage{subfig}

\begin{document}

\newcommand{\va}{\mathbf{a}}
\newcommand{\vb}{\mathbf{b}}
\renewcommand{\vr}{\mathbf{r}}
\newcommand{\vR}{\mathbf{R}}
\newcommand{\vK}{\mathbf{K}}
\newcommand{\pll}{\parallel}

\title{Tunable Luttinger liquid physics in biased bilayer graphene}

\author{Matthew Killi}
\affiliation{Department of Physics, University of Toronto, Toronto, Ontario, M5S1A7, Canada}

\author{Tzu-Chieh Wei}
\affiliation{Department of Physics and Astronomy, University of British Columbia, Vancouver, BC, Canada V6T1Z1}

\author{Ian Affleck}
\affiliation{Department of Physics and Astronomy, University of British Columbia, Vancouver, BC, Canada V6T1Z1}
\affiliation{Canadian Institute for Advanced Research, Toronto, Ontario, M5G 1Z8, Canada}

\author{Arun Paramekanti}
\affiliation{Department of Physics, University of Toronto, Toronto, Ontario, M5S1A7, Canada}
\affiliation{Canadian Institute for Advanced Research, Toronto, Ontario, M5G 1Z8, Canada}

\date{\today}

\begin{abstract}
Electronically gated bilayer graphene behaves as a tunable gap semiconductor under a uniform interlayer bias $V_{g}$.  Imposing a spatially varying bias, which changes polarity from $-V_g$ to $+V_g$, leads to one dimensional (1D) chiral modes localized along the domain wall of the bias.  Due to the broad transverse spread of their low-energy wavefunctions, we find that the dominant interaction between these 1D electrons is the forward scattering part of the Coulomb repulsion. Incorporating these interactions and the gate voltage dependence of the dispersion and wavefunctions, we find that these 1D modes behave as a strongly interacting Tomonaga-Luttinger liquid with three distinct mode velocities and a bias dependent Luttinger parameter, and discuss its experimental signatures.
\end{abstract}
\maketitle
Graphene has been the focus of intense research in recent years \cite{Castro-Neto:2009qy} due to the rich physics
of its massless Dirac fermions.  
Theoretical proposals have shown that spatially modulating an applied gate voltage in monolayer graphene leads to 
tunable anisotropic transport properties \cite{Park:2008yq}. 
For bilayer graphene (BLG), it was experimentally demonstrated that applying an electric field perpendicular to the layers, by sandwiching it between two gates, induces a gap in its electronic spectrum that is of the order of the applied interlayer bias \cite{San-Jose:2009,Castro:2007yq,Oostinga:2008fk,Zhang:2009ys}. Biased BLG thus behaves as a semiconductor with a tunable band gap. Such tunability of electronic properties opens up possibilities for
graphene based devices \cite{Liu:2008lr}.

\begin{figure}[b]
\subfloat[]{
\label{schematic} 
\includegraphics[width=4.2cm]{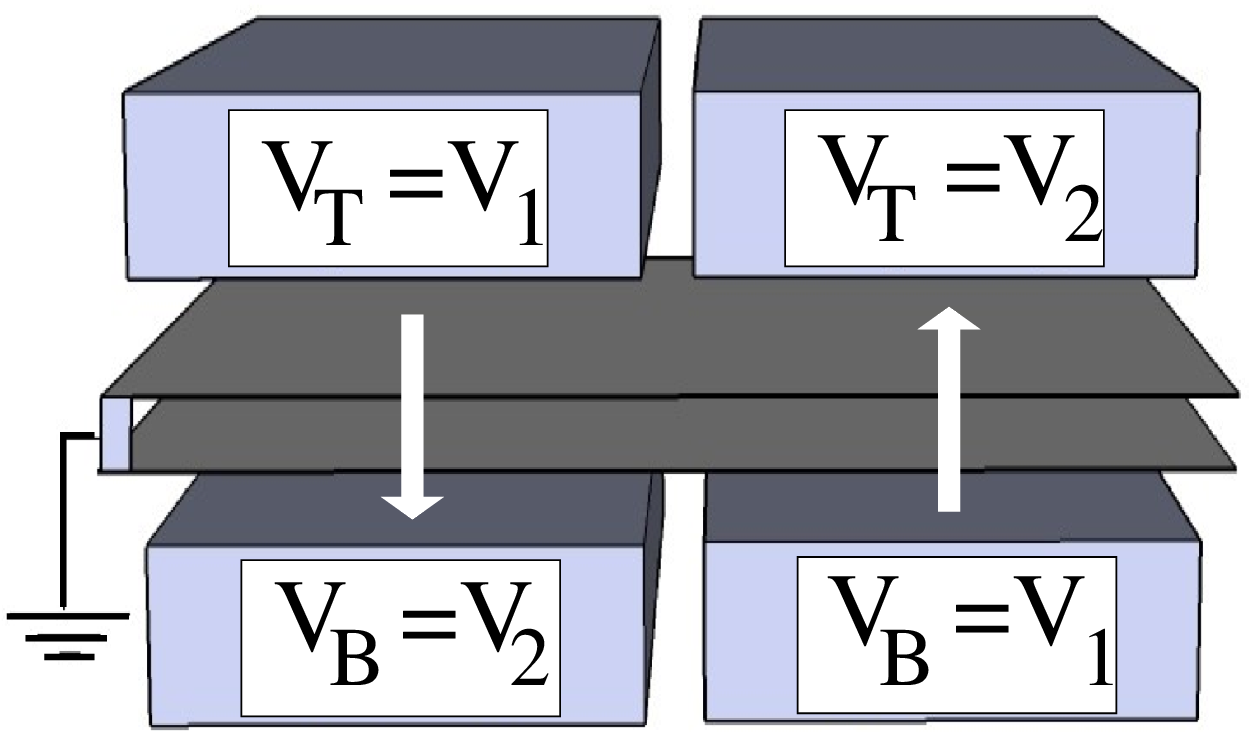}}
\hspace{-.2cm}
\subfloat[]{ 
\label{lattice} 
\includegraphics[width=4.2cm]{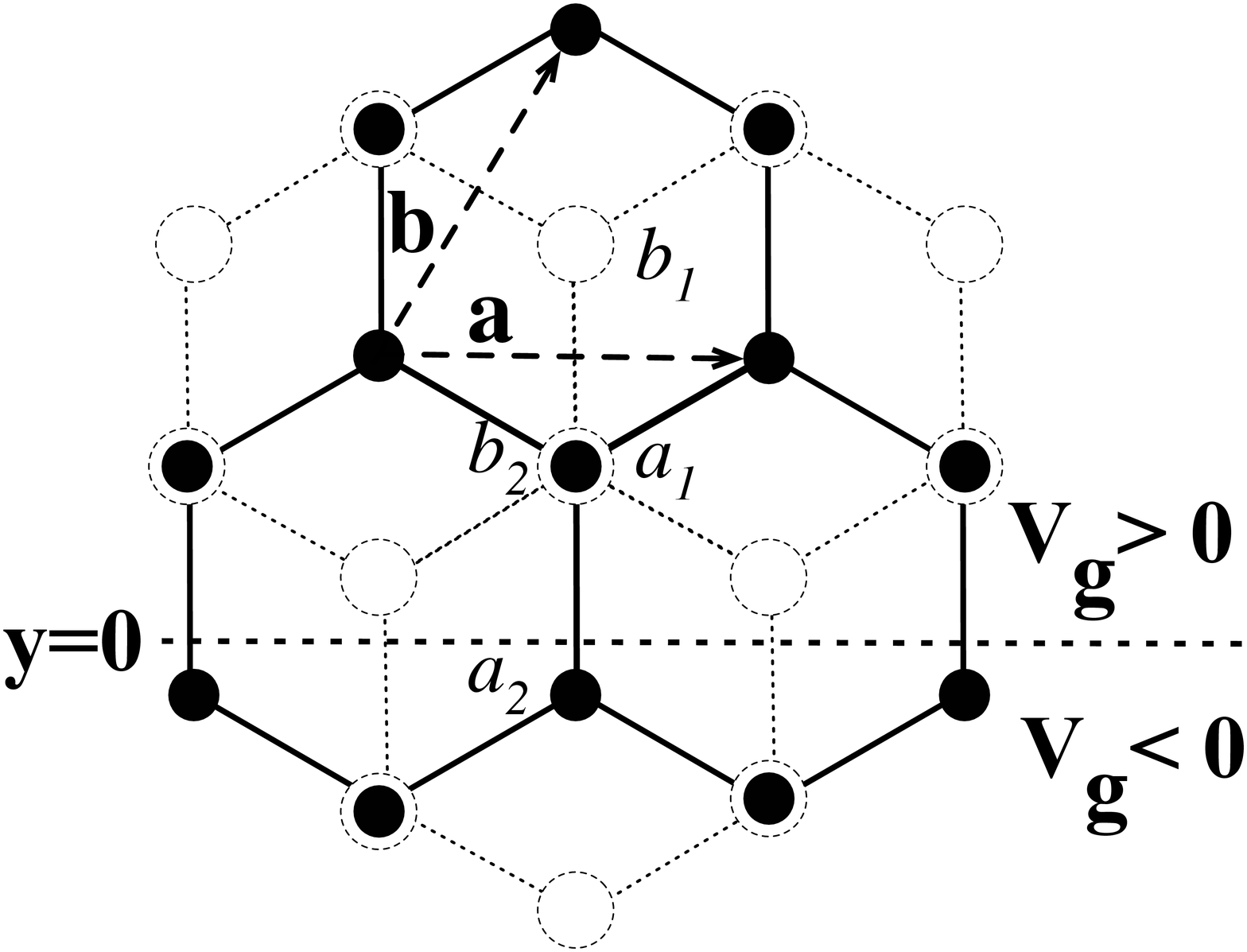}}
\caption{ (a) Schematic diagram of external gates. $V_{\rm T}+V_{\rm B}$ controls the bilayer doping while
$V_{\rm T}-V_{\rm B}$ controls the gap via the depicted perpendicular electric field.  (b) Structure of bilayer graphene with 
a `bias kink' in $V_g=V_{\rm T}-V_{\rm B}$ at $y=0$.}
\vspace{-0.5cm}
\end{figure}

Recently, Martin, Blanter and Morpurgo \cite{Martin:2008kx} have proposed a setup 
where two neighbouring regions of BLG are biased with opposite parity, shown 
schematically in Fig.\ (\ref{schematic}).  In this geometry, gapless one dimensional (1D) modes were 
shown to emerge at the interface where the bias reverses sign (see also Ref.\cite{Yao:2009uq}).  These 
modes are analogous to
domain wall fermions studied in the context of polyacetylene \cite{Heeger:1988rt}, charge density waves in
graphene \cite{Semenoff:2008vn}, and field theories in high-energy physics \cite{Jackiw:1976kx}. 
These modes in BLG may also be 
viewed as switchable nanowires which can
be turned on and off using different gate voltages for a given gate
configuration.

In this Letter, we study the effects of interactions on the low-energy modes of a single such wire in the original setup proposed in Ref.\cite{Martin:2008kx}.  
We find that the low-energy wavefunctions of the 1D modes have a broad spread in the direction transverse to the interface --- this leads to the dominance of the forward scattering part of the Coulomb interaction between electrons, 
in a manner akin to large radius carbon nanotubes \cite{Kane:1997zl}. 
Within an abelian bosonization framework, incorporating these forward scattering terms is shown to lead to a strongly interacting Tomonaga-Luttinger liquid \cite{Giamarchi:2004fr}.
Remarkably, we find the Luttinger parameter of this liquid is tunable by adjusting the gate potential. This results from two competing effects: 
(i) An increased bias causes further confinement of the wavefunctions to the interface, enhancing the effect of interactions; (ii) An increase in the bias increases the Fermi velocity of the low-energy modes, suppressing the effect of interactions relative to the kinetic energy. The net result is that the Luttinger parameter in the total charge channel,
$K_{c +}$, can be varied between $0.15$-$0.2$ by increasing the bias over an experimentally accessible range. At the
same time the Luttinger parameter in the transverse charge channel, $K_{c-} \approx 0.63$, is relatively independent of the bias. We thus show that gated BLG can realize a tunable Tomonaga-Luttinger liquid. 
Such band structure and wavefunction tuning of Luttinger liquids
has been suggested in a few other systems recently --- in cold atomic gases \cite{Zhai:2005vn, Moritz:2005kx}, in magnetic waveguides in graphene \cite{Hausler:2008rt}, in carbon nanotubes in crossed electric and magnetic fields \cite{DeGottardi:2009uq,DeGottardi:2009vn},
and in gated topological insulators \cite{Yokoyama:2010fk}.
Such
a Luttinger liquid with dominant forward scattering is also expected to arise at charge density wave (CDW)
domain walls in single layer graphene, where the CDW involves a weak 
sublattice density modulation induced by an appropriate substrate
\cite{Semenoff:2008vn}. 

\textit{Non-Interacting System.} --- 
BLG consists of two Bernal stacked graphene layers as depicted in Fig.~\ref{lattice}.  We label the carbon atoms in the bilayer by a unit cell index $i$, a sub-lattice index $s=a,b$, and a layer index $\ell=1,2$ labelling top and bottom layers respectively.  The distance between neighboring carbon atoms in the same layer and on the same sublattice is $d \approx 2.46 \AA$, while the interlayer distance is $d_\perp \approx 3.34 \AA$.  
The minimal tight-binding model for electrons in BLG consists of a nearest-neighbor hopping amplitude $t \approx 3 eV$ within each layer, and an interlayer hopping amplitude $t_\perp \approx 0.15 eV$ between sites 
$(i,s\!=\!a,\ell\!=\!1)$ and $(i,s\!=\!b,\ell\!=\!2)$.
Henceforth, we use units where $\hbar\!=\!t\!=\!d\!=\!1$, and 
set $i \equiv m \va + n \vb$ where $\va,\vb$ are unit vectors depicted in Fig.~\ref{lattice}.

\begin{figure}[t]
\subfloat[]{
\label{V0.01} 
\includegraphics[width=3.1cm, angle=-90]{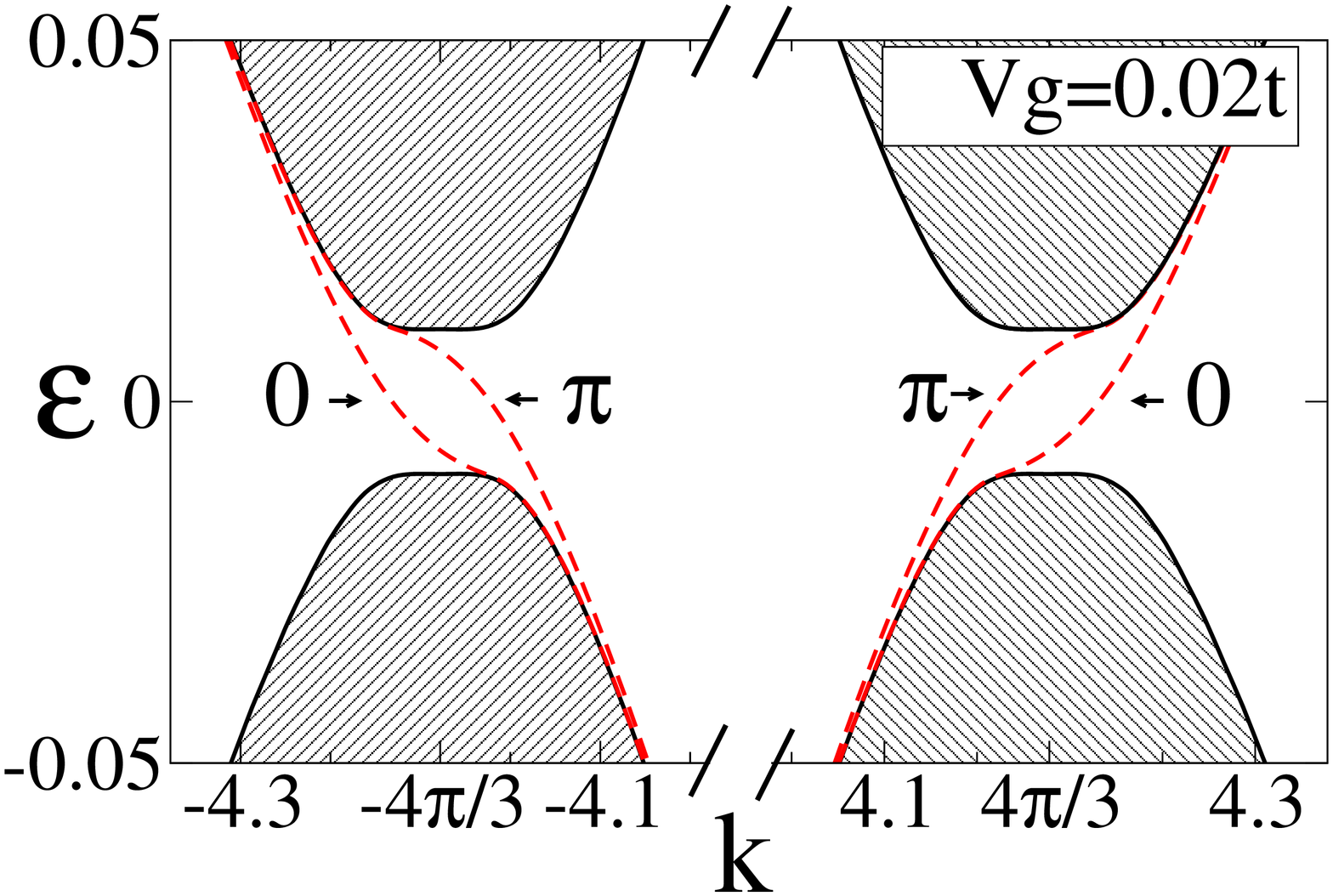}}
\hspace{-.43cm}
\subfloat[]{ 
\label{V0.40} 
\includegraphics[width=3.1cm, angle=-90]{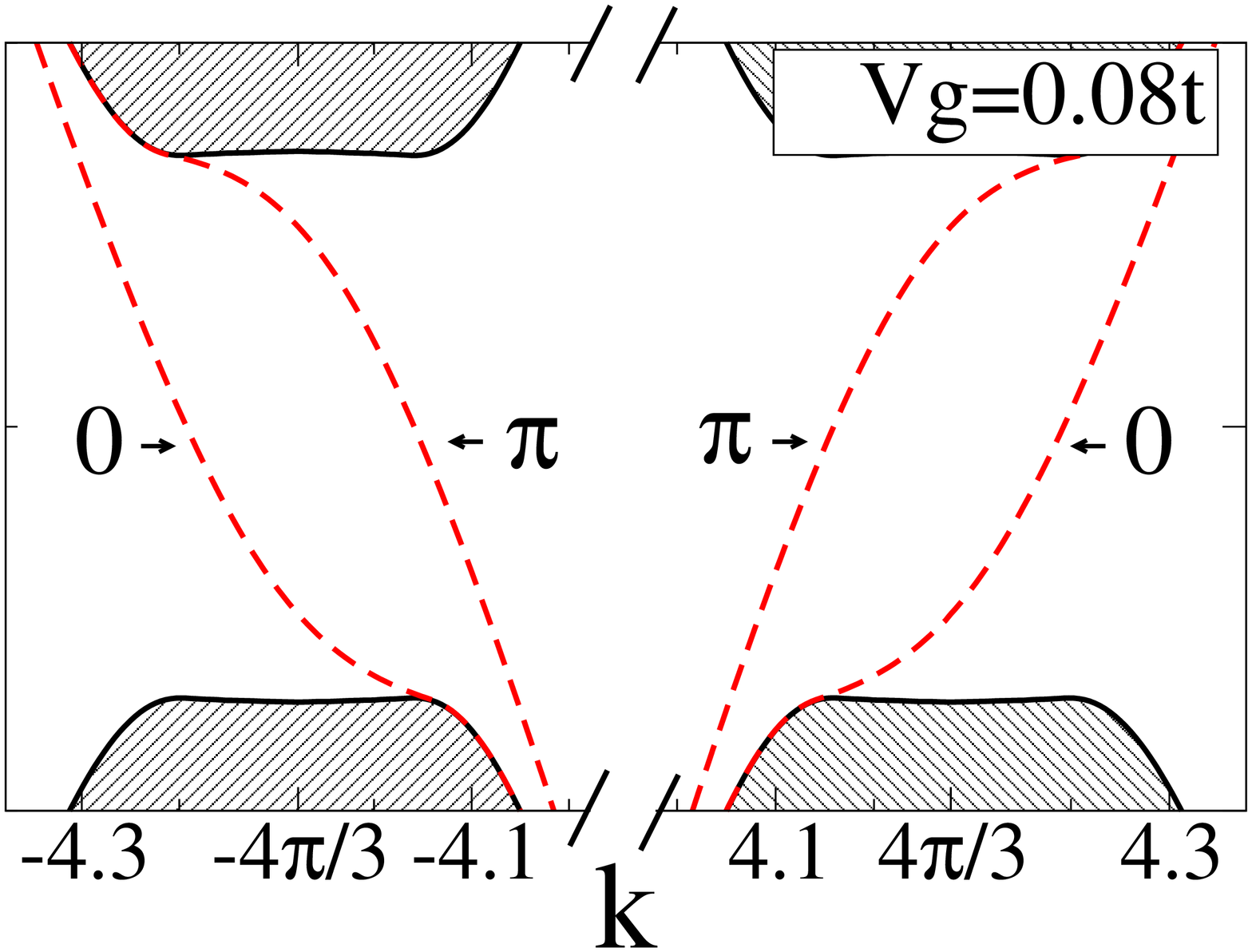}}
\caption{Dispersion about the K-points with (a) $V_{g}=0.02t$ and (b) $V_{g}=0.08t$.  Edge-mode bands are indicated by the labelled arrows and bulk-states by the hatched region.}
\label{Dispersions}
\end{figure}

In order to describe the effect of the external gates, we add a potential term,
$-\frac{1}{2} \! \sum_{\vR,\sigma}\!\!  \left(-1\right)^{\ell}V_g (y_{n,s,\ell}) \, \hat{n}_{\vR,\sigma}$,
to the tight-binding Hamiltonian, where $\hat{n}_{\vR,\sigma}$ is the electron number operator for spin-$\sigma$ at the site labeled by the set of indices $\vR=(m,n,s,\ell)$. Here, we have assumed that the bias $V_g(y)$ depends only on the $y$-coordinate, which is determined by $(n,s,\ell)$, and is $m$-independent so that translational symmetry is preserved along $\va$ \cite{footnote1}.  We assume $V_T\!+\!V_B$ (see Fig.~1) is chosen to fix the bilayer chemical potential at the mid-gap.
For later convenience, we set $\vR=(m,\vr)$, where $\vr\equiv(n,s,\ell)$.  For a general potential profile with $V_g(y>0)=-V_g(y<0)$ and $V_g( y\to\pm\infty) = \pm V_g$, the bulk region far from $y=0$ has a gap $\Delta \approx V_g$, while the interface has been shown to support gapless 1D modes \cite{Martin:2008kx}. Here, for simplicity, we assume the gates induce a potential with a step profile $V_g(y)=V_g {\rm sign}(y)$, since the dominant effect of a potential can be shown to come from the strength of the bias and not the details of its spatial profile \cite{footnote2}. 
This Hamiltonian can be solved by Fourier transforming along the direction parallel to the interface, with momentum labelled $k$, and then numerically diagonalizing the matrix for each $k$.  Our numerical results for various gate biases that have been shown to be  experimentally viable \cite{Zhang:2009ys}, (see Fig.\ (\ref{Dispersions})) show two right (left) movers for each spin at the $K$ (-$K$) point, consistent with Ref.~\cite{Martin:2008kx}. Drawing an analogy with two leg Hubbard ladders \cite{Fabrizio:1992ye,Balents:1996to,Chudzinski:2008}, 
we refer to the high/low energy bands as the $\pi$/$0$-bands.

\begin{figure}[t]
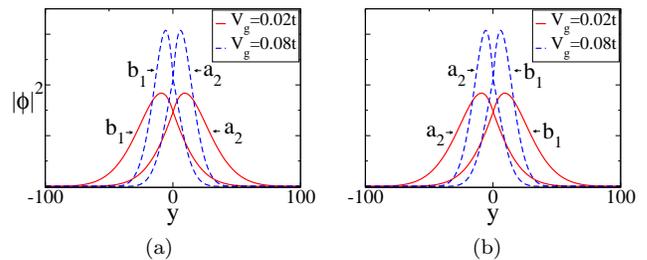

\subfloat[]{
\label{WFO} 
\includegraphics[height=2.9cm]{WFO.eps}}
\hspace{0.2cm}
\subfloat[]{ 
\label{WFP} 
\includegraphics[height=2.9cm]{WFP.eps}}
\caption{The modulus square of the zero-energy wavefunction of (a) the $0$-band and (b) the $\pi$-band on the dominate sites $a_2$ and $b_1$ at $V_g=0.02t$ and $V_g=0.08t$ (independent of being a right or left mover).}
\label{GateSpread}
\end{figure}

As seen qualitatively from the figure, and explicitly in Fig.~(\ref{parameters}), the Fermi velocity of the two bands are equal and they both change significantly with the bias voltage.  Fig.\ (\ref{GateSpread}) shows the plots of the modulus square of the zero-energy wavefunctions in the $0$-band and $\pi$-band at high and low gate voltage.  We see from here that the spread of the wavefunction transverse to the interface direction varies significantly.  Note that the zero-energy wavefunction of the $\pi$-band is related by interchanging the layers and reflecting about the interface so that wavefunction leans in to the region of high potential instead. Both, this fact and the equality of the Fermi velocities of the two bands ($V_{F}^{0/\pi}=V_{F}$), are a consequence of the symmetry that relates the two band by an inversion about the $K$ (-$K$) point at low energies \cite{Martin:2008kx}.  

\textit{Effective 1D Hamiltonian.} --- 
To derive the effective low-energy 1D hamiltonian, we assume a suitable energy cutoff that is smaller than the bulk gap and focus on those single particle states that lie within this energy window and are confined to the 1D interface region.  To do this, we first expand the field operators in the complete basis, $\hat{\Psi}_{\vR,\sigma} = \frac{1}{\sqrt{L}} \sum_{k, \alpha}  e^{i k m d}\varphi^{\alpha}_{k}(\vr) \, \hat{\psi}^{\alpha}_{k \sigma}$, where $\varphi^{\alpha}_{k}(\vr)$ is the wavefunction of the state in band-$\alpha$ with momentum $k$.  We then restrict the bands to the set $\alpha=\{0,\pi\}$ and consider only momenta in the vicinity of the four Fermi points, $\pm k^{\pi}_{F}$ and $\pm k^{0}_{F}$.  An additional simplification is made by neglecting the small momentum dependence of the wavefunctions since $\varphi^{\alpha}_{\pm k^{\alpha}_{F}+q}(\vr) \approx \varphi^{\alpha}_{R/L}(\vr)$ for small momenta $q$, where $\varphi^{\alpha}_{R/L}\!(\vr)$ is the zero-energy wavefunction at $\pm k_{F}^{\alpha}$.  In doing so, the $\vr$-dependence of the wavefunction can be separated to yield the low-energy
field operators projected to the 1D subspace via
\begin{equation}
\label{field}
\hat{\Psi}_{\vR,\sigma} \approx
\sum_{r=\pm, \alpha=\{0,\pi\}} \varphi^{\alpha}_{r}(\vr) \, e^{i r k^{\alpha}_{F} x}  \hat{\psi}^{\alpha}_{r \sigma}  (x),
\end{equation}
where $r$ is the label R/L for left/right movers and takes the values $+/-$ in the expression, and $\psi^{\alpha}_{r \sigma}(x)$ are slowly varying field operators exclusively dependent on the position along the interface (which we now denote
by the continuous variable $x=m d$).

\begin{figure}[t]
\centering
\includegraphics[width=2.5cm]{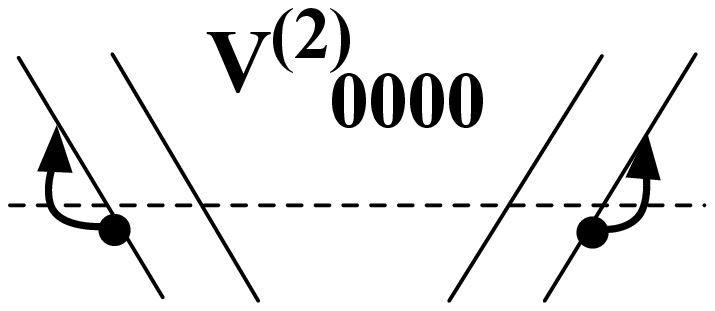}\hspace{0cm}
\includegraphics[width=2.5cm]{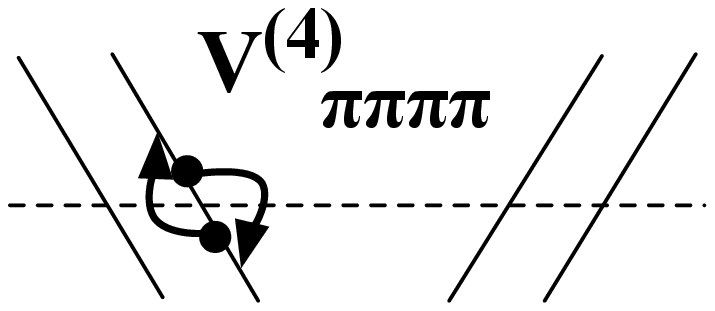}\hspace{0cm}
\includegraphics[width=2.5cm]{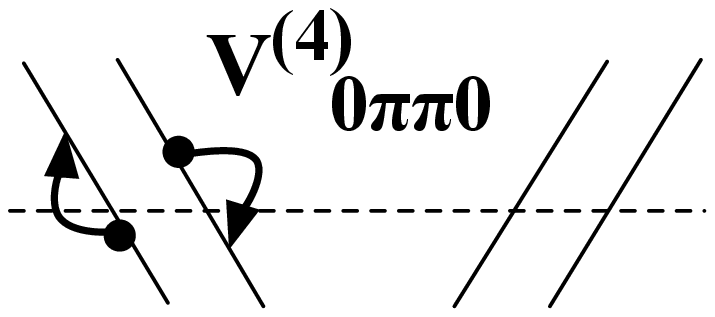}\\ \vspace{-.2cm}
\includegraphics[width=2.5cm]{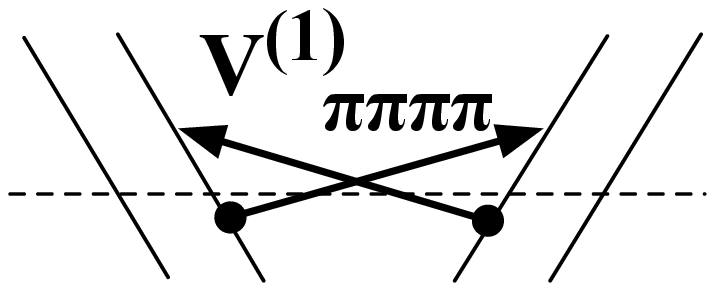}\hspace{0cm}
\includegraphics[width=2.5cm]{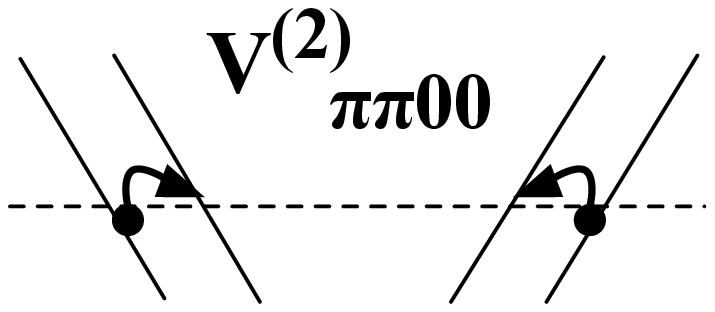}\hspace{0cm}
\includegraphics[width=2.5cm]{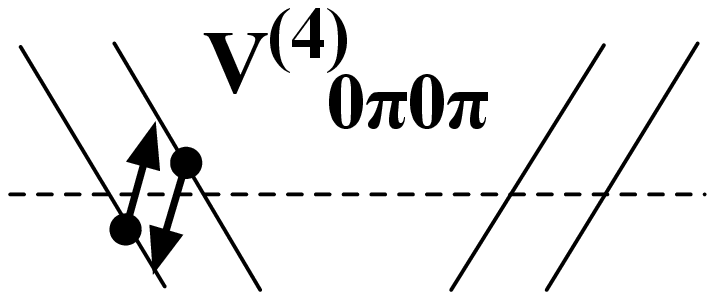}
\caption{\label{scattering}Examples of various scattering processes.}
\end{figure}

\begin{figure}[t]
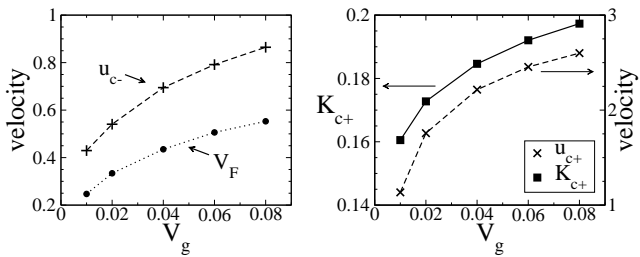

\includegraphics[height=3.3cm]{vf.eps}\hspace{0.2cm}
\includegraphics[height=3.3cm]{LLP.eps}
\vspace{-0.4cm}
\caption{Fermi velocity $V_F$ and mode velocities $u_{c\pm}$ in the charge sector (in units of $td$), and the Luttinger parameter of total charge sector as a function of $V_{g}$.  We assume a screening length of $1000 d \approx 0.3\mu m$ for the Coulomb interaction and a short distance cutoff of 0.5d.}
\label{parameters}
\vspace{0 cm}
\end{figure}

We now rewrite the entire Hamiltonian in terms of operators in the reduced 1D subspace.  The free part is simply linearized to give $\sum_{|q|<\Lambda} \sum_{r \alpha \sigma} rq V_{F} \hat{\psi}^{\alpha\dag}_{r \sigma} (q) \hat{\psi}^\alpha_{r \sigma} (q)$.  The effective interaction between fermions in the 1D channel is obtained by a straightforward substitution of Eqn.\ (\ref{field}) into the Coulomb term, $\frac{1}{2}\sum_{\sigma \sigma '} \sum_{\mathbf{R R'}} \Psi^{\dag}_{\vR,\sigma} \Psi^{\dag}_{\vR',\sigma '} U(\vR,\vR') \Psi_{\vR',\sigma '}\Psi_{\vR,\sigma}$, followed by a summation over $\vr$. This gives rise to various scattering terms, many of which are rapidly oscillating and can be dropped.  The effective Hamiltonian obtained contains many terms of the general form
\begin{align}
\label{effective}
\frac{V^{(i)}_{\alpha \beta \gamma \delta}}{2}\sum_{x} 
\hat{\psi}^{\alpha \dag}_{r_{1} \sigma} \! (x)
\hat{\psi}^{\beta \dag}_{r_{2} \sigma'} \! (x)
\hat{\psi}^{\gamma}_{r_{3} \sigma'} \!(x)
\hat{\psi}^{\delta}_{r_{4} \sigma}  \!(x).
\end{align}
Here, $V^{(i)}_{\alpha \beta \gamma \delta} \equiv V^{(i)}_{\alpha \beta \gamma \delta}  (r_{1} k_{F}^{\alpha}-  r_{4}k_{F}^{\delta}\!)$ is the Fourier component of the effective 1D potential
\begin{align}
\tilde{V}^{(i)}_{\alpha \beta \gamma \delta}(x\!-\!x') \!=\! \! \sum_{\vr, \vr'} \!U\!(\vR,\vR') 
{\varphi^{\alpha}_{r_{1}}}^{\!*} \!(\vr)
{\varphi^{\beta}_{r_{2}}}^{\!*} \! (\vr')
{\varphi^{\gamma}_{r_{3}}} \! (\vr')
{\varphi^{\delta}_{r_{4}}} \! (\vr). \notag
\end{align}
The effective interaction Hamiltonian contains all terms of the form in Eqn.\ (\ref{effective}) that have a combination of 
$R/L$ and band indices that conserve (crystal) momentum.
The index $i$ classifies the scattering processes using standard g-ology notation \cite{Haldane:1981ys,Penc:1990fj,Sorella,Fabrizio:1992ye,Balents:1996to,Chudzinski:2008} (see Fig.\ (\ref{scattering})):  $i=1$ refers to backscattering, $i=2$ to forward-scattering involving both right and left movers, and $i=4$ to forward-scattering involving only right or only left movers.
The number of distinct processes is greatly reduced by the fact that 
all interband scattering terms with parallel spin merely renormalize the coefficients of a corresponding intraband term with parallel spin. 

This Hamiltonian is qualitatively similar to that obtained for Hubbard ladders \cite{Fabrizio:1992ye,Balents:1996to,Chudzinski:2008} with two significant
differences. First, both the wavefunctions and $V_{F}$ are sensitive to changes in the applied gate bias.  
The modified wavefunctions alter the distance dependence of the effective Coulomb interaction, while the change in $V_{F}$ adjusts the relative interaction strength parameterized by the fine-structure constant $\alpha \rightarrow \alpha \frac{c}{V_{F}}$.
Second, we note that the long-range nature of the Coulomb interactions, together with the large spread of the low-energy wavefunctions, causes the small momentum forward scattering processes to dominate. This is reminiscent of large radius single wall carbon nanotubes, where the extension of the wavefunctions around the tube radius suppresses the bare backscattering \cite{Kane:1997zl}. 
We have checked that the bare values of
these backscattering and interband scattering terms
are very small, consistent with this argument.
For instance, $V^{(1)}_{0000}/V^{(2)}_{0000} \sim 10^{-3}$ and $V^{(4)}_{0\pi0\pi}/V^{(2)}_{0000} \sim 10^{-2} $ at 
$V_g=0.02t$, so that such processes are expected to be important only at very low energy and temperature.
We therefore first focus on the forward scattering processes.

\textit{Bosonization.} --- Using the standard abelian bosonization procedure \cite{Giamarchi:2004fr}, we introduce the bosonic field $\hat{\phi}_{\alpha \sigma}(x)$ and the phase $\hat{\theta}_{\alpha \sigma}(x)$ whose spatial derivative $\partial_{x} \hat{\theta}_{\alpha \sigma} = \hat{\Pi}_{\alpha \sigma}$ is conjugate to $\hat{\phi}_{\alpha \sigma}(x)$.  Fermion operators can be represented in terms of these boson fields via $\hat{\psi}^{\alpha}_{r \sigma}(x) \sim e^{i \left( r \hat{\phi}_{\alpha \sigma}(x) - \hat{\theta}_{\alpha \sigma}(x) \right)}$.  It is a simple matter to rewrite the density-density interactions in the boson representation by means of  the relations $\partial_{x}  \hat{\phi}_{\alpha \sigma} = - \pi \left( \hat{\rho}_{R \alpha \sigma} + \hat{\rho}_{L \alpha \sigma}  \right) $ and $\partial_{x}  \hat{\theta}_{\alpha \sigma} =  \pi \left( \hat{\rho}_{R \alpha \sigma} - \hat{\rho}_{L \alpha \sigma}  \right)$.  In addition, the symmetry between the bands allows us to lighten our notation by
defining $V_{A}\!\equiv \!V^{(2)}_{\alpha \alpha \alpha \alpha}\!=\!V^{(4)}_{\alpha \alpha \alpha \alpha}$ and $V_B \! \equiv \! V^{(2)}_{\alpha \bar{\alpha} \bar{\alpha} \alpha} \!=\!V^{(4)}_{\alpha \bar{\alpha} \bar{\alpha} \alpha}$.

This leads to the Hamiltonian,
\begin{align}
H_{1}=
\frac{1}{2 \pi} \int \! dx (\partial_{x} \Phi)^{T} \hat{u} \cdot \hat{K}^{-1} (\partial_{x} \Phi) + (\partial_{x} \Theta)^{T} \hat{u} \cdot \hat{K} (\partial_{x}  \Theta),
\end{align}
with
\begin{align}
\hat{u} \cdot \hat{K}^{-1} & = V_{F} \mathbf{1} +\frac{V_{F}}{2 \pi}
\begin{pmatrix}
      g_{A}	 		&	g_{B}			& 	g_{A}	&	g_{B} \\
      g_{B} 			&  	g_{A}			&	g_{B}	&	g_{A} \\
      g_{A}			&	g_{B}			&	g_{A}	&	g_{B} \\
      g_{B}			&	g_{A}			&	g_{B}	&  	g_{A}
\end{pmatrix} \\
\hat{u} \cdot \hat{K} & = V_{F} \mathbf{1}.
\end{align}
Here $g_{A/B} \equiv (2V_{A/B})/ V_{F}$, and $\Phi=(\phi_{0 \uparrow},   \phi_{\pi \uparrow}, \phi_{0 \downarrow}, \phi_{\pi \downarrow})^{T}$ with a similar definition for $\Theta$.

This Hamiltonian is diagonal in the total/transverse basis defined via 
$\phi_{\nu \pm}=\phi_{\nu 0} \pm \phi_{\nu \pi}$, where $\nu$ labels the spin ($s$) or charge ($c$) sector and $\phi_{c(s) \alpha}=\phi_{\alpha \uparrow} \pm \phi_{\alpha \downarrow}$.
In this basis, the spin and charge sectors decouple. The spin modes are unaffected by
interactions, the Luttinger parameters $K_{s \pm} \!=\! 1$ and the velocities
$u_{s \pm} \!= \!V_F$.  The charge modes have renormalized velocities and
nontrivial Luttinger parameters, given by
\begin{align}
u_{c \pm} & = V_{F} \left(1+ y_{c \pm}\right)^{\frac{1}{2}} \\
K_{c \pm} &= \left (1+y_{c \pm} \right)^{-\frac{1}{2}},
\end{align}
where $y_{c \pm} = 2 (V_A \pm V_B)/ \pi V_{F} $.  At the Gaussian level, the only effect of the interactions is thus 
to strongly modify $K_{c \pm}, u_{c \pm}$.
Fig.(\ref{parameters}) shows these parameters plotted for various gate voltages.
As seen from the figure, $K_{c+}$ can
be tuned significantly by the external bias; by contrast, $K_{c-} \approx 0.63$ (not shown) is relatively bias independent.

\textit{Observable consequences.} --- 
The strong interactions in the charge channel lead to three different velocities for the spin ($V_F$) and charge ($u_{c\pm}$)
modes in the
Luttinger liquid. Mapping out these dispersing modes, as has been done in semiconductor heterostructures
\cite{Auslaender:2005},
appears to be challenging in the biased BLG system. A more accessible
signature of the Luttinger liquid physics is the energy dependence of the single particle density of states (DOS). We
expect $n(\epsilon) \sim \epsilon^{\alpha}$, with $\alpha > 0$.
Such a suppression of the DOS is expected to lead to a tunneling conductance 
$G \sim T^\alpha$ (for voltages $e V \! \ll \! k_B T$)
or a nonlinear differential conductance $dI/dV \sim V^\alpha$ (for $e V \! \gg \! k_B T$). We find
$\alpha_{\rm bulk} = \frac{1}{8} (K_{c+}+K_{c+}^{-1}+K_{c-}+K_{c-}^{-1} - 4)$ and
$\alpha_{\rm edge} = \frac{1}{4}(K_{c+}^{-1}+K_{c-}^{-1} - 2)$,
so that bias dependent tunneling exponents are expected to be observed.
Various charge density, spin density and superconducting pair
correlators are expected to show power-law decays in this intermediate energy Luttinger liquid regime.
We find that charge and spin density wave operators at $2 k_F^0$ and $2 k_F^\pi$ are most strongly
enhanced by interactions in this regime, decaying along the interface as $|x-x'|^{-(2+\!K_{c+}\!+\!K_{c-}\!)/2}$, with
the precise lattice scale modulation pattern in $(n,s,\ell)$
being determined by the prefactors set by the wavefunctions $\varphi^{\alpha}_{R/L}(\vr)$ from Eq.(\ref{field}).
Modulations at $k_F^0\!\pm\!k_F^\pi$ are subdominant.

\textit{Discussion.} --- We have shown that BLG in a suitable gate geometry can
realize a tunable Luttinger liquid with four gapless modes. 
This is expected to break down once
backscattering and interband scattering terms become important; since the bare values of these
interactions are small and that
they are all marginal, and thus flow slowly, there is an intermediate energy
window where the physics discussed above should be observable. At very low energy (or
temperature), however, we expect that such processes will gap out all 
sectors except the $c+$ channel,
which should still exhibit Luttinger liquid physics.
Turning to the effect of additional
interlayer hopping terms ($\gamma_3$) between the $a_2$ and $b_1$ sites, we have checked that
this renders $V_F^0 \neq V_F^\pi$.
This asymmetry is small for moderate bias voltages; further, interband
scattering tends to equalize the velocities \cite{Penc:1990fj,Chudzinski:2008}
so that small velocity asymmetries are expected to be unimportant. We have assumed that the Fermi level is
tuned, via $V_T+V_B$, to be
precisely in the middle of the gap - small deviations which tend to slightly
dope the interface states while leaving the
bulk gap intact will not qualitatively alter the physics discussed here.
While the `domain wall' modes discussed here are topologically protected \cite{Martin:2008kx} independent of the
precise bias profile transverse to the wire, disorder along the wire direction
will lead to backscattering. Backscattering is somewhat mitigated by the wavefunction spread, but
we expect it will lead to insulating behavior at very low energy \cite{KaneFisher:1992}. Fourier transform
scanning tunneling 
spectroscopy would then be useful to uncover the underlying Luttinger liquid physics \cite{Kivelson:2003}.

\acknowledgments
We acknowledge useful discussions with 
K. Burch, P. Chudzinski, C. N. Lau, and D. Podolsky. This work was 
supported by NSERC  of Canada (MK,TCW,IA,AP), MITACS (TCW), the Sloan Foundation (AP), and
an Ontario ERA (AP).


\begin{thebibliography}{999}
\bibitem{footnote1}
A slight change in the interface direction changes the projection of the K-points along the translation-invariant direction and modifies the Fermi points but does not lead to a qualitative change in the physics as long as the interface is not 
aligned close to an armchair direction.
\end{thebibliography}
\end{document}